\documentstyle[12pt,aasms4]{article}
\newcommand{\FLUXUNIT}{${\rm erg}\ {\rm cm}^{-2}\ {\rm s}^{-1}$}
\newcommand{\NHUNIT}{${\rm H}\ {\rm cm}^{-2}$}

\newcommand{\etal}{et al.}
\newcommand{\ASCA}{{\it ASCA}}

\newcommand{\ROSAT}{{\it ROSAT}}

\received{1997, Dec 15}
\accepted{1998, Apr 1}
\lefthead{Sakano et al.}
\righthead{The hardest X-ray source in the ASCA LSS}

\begin{document}

\title{The hardest X-ray source in the \ASCA\ Large Sky Survey:\\
Discovery of a new type 2 Seyfert}

\author{Masaaki Sakano\altaffilmark{1}, Katsuji Koyama\altaffilmark{2}, Takeshi Tsuru\altaffilmark{2} and Hisamitsu Awaki\altaffilmark{2}}
\affil{Department of Physics, Kyoto University,
 Kyoto 606-8502 Japan,
 sakano@cr.scphys.kyoto-u.ac.jp, koyama@cr.scphys.kyoto-u.ac.jp,
 tsuru@cr.scphys.kyoto-u.ac.jp, awaki@cr.scphys.kyoto-u.ac.jp}

\author{Yoshihiro Ueda and Tadayuki Takahashi}
\affil{Institute of Space and Astronautical Science,
 Kanagawa 229-8510 Japan,
 ueda@astro.isas.ac.jp, takahasi@astro.isas.ac.jp}

\author{Masayuki Akiyama\altaffilmark{1} and Kouji Ohta}
\affil{Department of Astronomy, Kyoto University, Kyoto 606-8502
    Japan,
 akiyama@kusastro.kyoto-u.ac.jp, ohta@kusastro.kyoto-u.ac.jp}

\and

\author{Toru Yamada}
\affil{Astronomical Institute, Tohoku University,
     Sendai 980-8578 Japan,
 yamada@astr.tohoku.ac.jp}

\altaffiltext{1}{Research Fellow of the Japan Society for the Promotion 
of Science.}
\altaffiltext{2}{CREST: Japan Science and Technology Corporation (JST)}

\begin{abstract}

We present results of \ASCA\ deep exposure observations of the hardest
X-ray source discovered in the \ASCA\ Large Sky Survey (LSS) project,
designated as AX~J131501$+$3141. We extract its accurate X-ray
spectrum, taking account of the contamination from a nearby soft
source (AX~J131502$+$3142), separated only by 1$'$.
AX~J131501$+$3141 exhibits a large absorption of $N_{\rm H} =
(6^{+4}_{-2})\times 10^{22}$ \NHUNIT\ with a photon index $\Gamma =
1.5^{+0.7}_{-0.6}$.  The 2--10 keV flux was about $5\times
10^{-13}$ \FLUXUNIT\ and was time variable by a factor of 30\% in 0.5
year.  From the highly absorbed X-ray spectrum and the time variability,
as well as the results of the optical follow-up observations
(\cite{Akiyama98}), we conclude that AX~J131501$+$3141 is a type 2
Seyfert galaxy.  Discovery of such a low flux and highly absorbed
X-ray source could have a significant impact on the origin of the
cosmic X-ray background.
\end{abstract}


\section{Introduction}

Since the discovery of the Cosmic X-Ray Background (CXB) more than 30
years ago (\cite{Giacconi62}), its origin has been a long standing
puzzle.  With the \ROSAT, $\sim$70\% of the CXB below 2 keV has been
resolved into point sources, more than 60\% of which are type 1 active
galactic nuclei (AGNs) (\cite{Vikhlinin95}; \cite{Hasinger96};
\cite{McHardy98}; \cite{Hasinger98}).  Origin of the CXB above the 2 keV band, however,
is less clear due to the absence of the imaging instrument in this
energy band.  One problem in the hard X-ray band, often referred as
the spectral paradox (e.g. \cite{Fabian92}), is that the X-ray
spectrum of the CXB in the 2--10 keV band is harder than that of the
typical type 1 AGN, which is presumably the main contributor to the
CXB below 2 keV.  The 2--10 keV X-ray spectra of type 1
Seyfert galaxies (most of the bright AGNs) are approximated by a
power-law with a mean photon index of 1.7 (\cite{Mushotzky93}), which
is significantly steeper than that of the 2--10 keV CXB of about 1.4
(\cite{Gendreau95}).  This fact implies that the origin of the CXB
above 2 keV differs, at least in part, from that below 2 keV.  In
addition, 20\% of the total energy of the CXB is contained in the 2--10 keV band,
whereas only a few percent of the CXB is contained below 2 keV (see review
by Fabian \& Barcons (1992), Hasinger (1996)).  Thus, the 2--10 keV
band would be an essential energy range to solve the origin of the CXB.

\ASCA\ is the first satellite with the capability of the hard
X-ray (up to 10keV) imaging and spectroscopy, hence is presently the
best satellite to investigate the CXB in the 2--10 keV band.  In the
\ASCA\ Large Sky Survey project (LSS: \cite{Inoue96}; \cite{Ueda98a}),
a continuous field of 7 deg$^2$ near the north galactic pole was
surveyed with a sensitivity higher than any previous surveys in this
energy band.  Ueda \etal\ (1998a) resolved a significant fraction of
the CXB, about 30\% of the CXB, into discrete sources at a sensitivity
limit of $F_{\rm X}\sim 10^{-13}$ \FLUXUNIT\ (2--10 keV).  The mean
photon index in the 2--10 keV band for these resolved sources ($F_{\rm
X}=$(1--4)$\times 10^{-13}$ \FLUXUNIT\ in 2--10 keV) was found to be
$\Gamma = 1.5\pm 0.2$.  This result is consistent with the idea that
the photon index approaches to that of the CXB,
$\Gamma\sim 1.4$, as the source flux decreases.
However, due to
limited photon statistics, the spectral information of the resolved
sources was too poor to address the nature of individual X-ray sources.

The hardest source in the LSS (hereafter we refer it as the ``LSS
hardest source'') was found to show a photon index of $\Gamma\sim -0.2$
with no correction of an absorption (\cite{Ueda96}). However, it is
unclear whether the apparent hard spectrum is due to a large
absorption or due to flatness of the intrinsic spectrum.  The LSS hardest
source, which was found in an unbiased survey, would provide us a good
opportunity to investigate the nature of faint and hard sources which
could significantly contribute to the CXB above 2 keV. Hence, we have
performed follow-up \ASCA\ and optical observations on the LSS hardest
source. This paper reports results of the \ASCA\ deep exposure
observations, while those of the optical observations are given by
Akiyama \etal\ (1998).

\section{Observations and Data Reductions}

We have made two follow-up \ASCA\ observations of the LSS hardest source: 
 on 1995 December 23--24 for approximately 51,000 seconds and on
1996 June 6--8 for approximately 49,000 seconds.  \ASCA\ has
four X-ray telescopes (XRT) with focal plane detectors of two Solid
State Imaging Spectrometers (SIS0 and 1) and two Gas Imaging
Spectrometers (GIS2 and 3).  Details of these instrumentation are
found in Serlemitsos \etal\ (1995), Burke \etal\ (1991, 1994), Ohashi
\etal\ (1996), Makishima \etal\ (1996), while a general description of
\ASCA\ can be found in Tanaka, Inoue, \& Holt (1994).  The observation
modes were 1-CCD FAINT mode and PH nominal mode for SISs and GISs,
respectively.  Data reduction and cleaning were made with the standard
method (\cite{Day95}).

\section{Results and Analysis}

% \placefigure{fig:image}

\subsection{X-ray Images \label{sec:image}}

\begin{figure}
% \centering
% \epsfile{file=fig/fig1.eps,width=\textwidth}
% \figcaption[fig1.eps]{
\caption[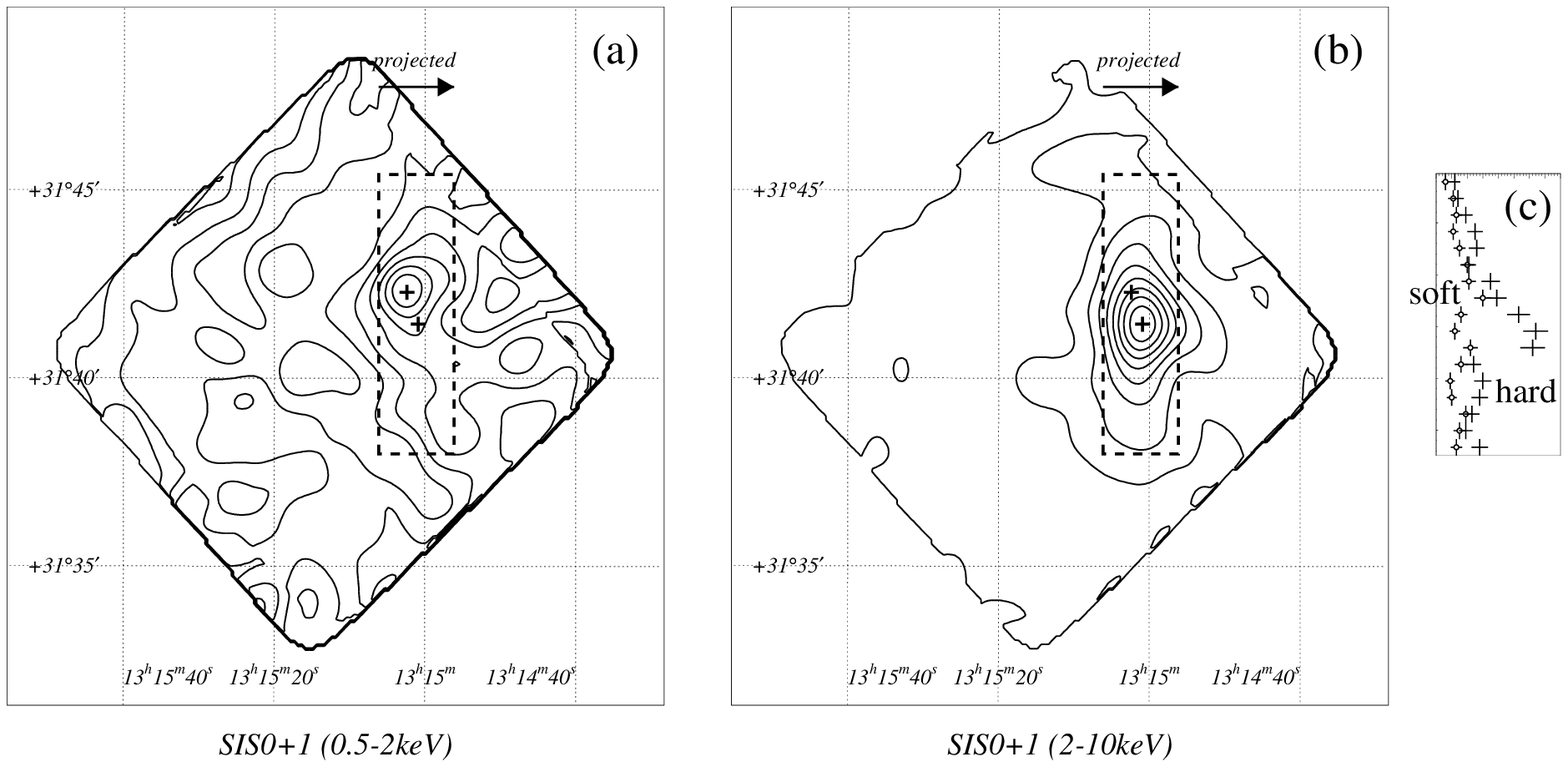]{
(a, b) SIS0$+$1 image contours in the 2nd observation  (1996 June).
The coordinates are in J2000.  Images are smoothed with Gaussian
of $\sigma=$ 20 pixels ($\sim$ 0.54 arcmin.).
The contour levels are linearly spaced by ten lines
from 0 to $2.0\times 10^{-3}$ c/s/16pixels and 
from 0 to $5.2\times 10^{-3}$ c/s/16pixels,
 for (a) the 0.5--2 keV and (b) the 2--10 keV band image,
respectively.
  The peak positions corresponding to AX~J131502$+$3142 and AX~J131501$+$3141 are marked
with crosses.  (c) The projected profiles of the 0.5--2 keV and the 2--10 keV band
in the region represented with dashed lines.
 The data represented with open circles and crosses correspond to that for the
0.5--2 keV and the 2--10 keV band data, respectively. 
\label{fig:image}
}
\end{figure}

 The LSS hardest source was detected in both the observations. 
Fig.~\ref{fig:image} shows the SIS image obtained from the 2nd
observation (a) in the soft band (0.5--2 keV) and (b) in the hard band
(2--10 keV). The images are the sums of data from SIS0 and SIS1,
smoothed with a Gaussian of $\sigma =$ 20 pixels (0.54 arcmin).  One
point-like source is clearly seen in the 2--10 keV band image, which
corresponds to the LSS hardest source.  On the other hand, in the
0.5--2 keV band image, we detected another point-like source whose position
shifts by about 1$'$ north from the peak found in the 2--10 keV band image,
but found no significant emission at the position of the
LSS hardest source. The relative accuracy of the peak position of
\ASCA\ mainly depends on photon statistics.  The photon counts of the
sources were 120 counts and 380 counts in the 0.5--2 keV and the 2--10
keV band, respectively, and the relative accuracy is estimated to be
$\sim 15''$.  Hence, we conclude that the peak position in the 0.5--2
keV band corresponds to another new soft X-ray source (here, the soft source),
which was not found in the LSS due to lack of the photon
statistics. Both of the sources are marked with crosses in
Fig.~\ref{fig:image}~(a, b).

 The nominal error of 1$'$ for the absolute position of \ASCA,
is mainly affiliatable to a mis-alignment of the attitude sensors
on the satellite base plate, coupled to a mis-match of the thermal
expansion coefficient (\cite{Ueda98b}). We
restored this temperature-dependent error using the on board
house-keeping data, according to the method described in Ueda \etal\
(1998b).  Finally, the peak positions for the soft and the LSS hardest source
are determined to be (13h 15m 1.8s, $+$31$^{\circ}$ 42$'$ 20$''$) and (13h 15m 0.9s,
$+$31$^{\circ}$ 41$'$ 28$''$), respectively, in the 2000 equinox with error circles of 0.5$'$ radii
(90\% confidence level). Accordingly, they are designated as
AX~J131502$+$3142 and AX~J131501$+$3141.

\subsection{Spectra determined by the image analysis\label{sec:image-fit}}

Since the half power diameter of \ASCA\ point spread function (PSF) is
about 3$'$, it is difficult to make individual spectra of these two
sources separated by only 1$'$.  Hence, we extracted the energy
spectra by analyzing images using the method described in Uno (1997).
We made projected profiles of six different energy bands in the region
illustrated in Fig.~\ref{fig:image}~(a, b) (dashed lines).
Each energy band was selected to contain reasonable counts, at least 100 counts,
 for the profile
fitting analysis as described below.
  We only
used the data of SIS0 and SIS1 to construct the combined images,
because the angular resolution of the SISs is better than that of the
GISs. The projected axis was selected along a constant Right Ascension
line which gives roughly the largest separation angle between the two
sources.  The profiles in the 0.5--2 keV and the 2--10 keV band are
separately shown in Fig.~\ref{fig:image}~(c).  We fitted these
profiles with a model consisting of a background and two projected
PSFs. The PSFs of XRTs were constructed by a
ray-tracing program (\cite{Kunieda95}).  The systematic error in the
shapes of the PSFs is about 10\% (\cite{Kunieda95}), which is much
smaller than the statistical error. For the background, we used a
model consisting of the CXB and the Non X-ray Background (NXB). The
former was produced by the ray-tracing program and the latter was
modeled from the night Earth data (Ueda 1996). In the fitting, we
fixed the positions, but allowed the fluxes of the two sources to vary.
Also the background level was varied.  The best-fit fluxes
of the two sources in the different energy bands give
the energy spectra.

   Using the energy spectra derived from the image fitting,
we examined whether the spectral shapes were
different between two observations, and found probable variability in
the total flux for the AX~J131501$+$3141 but found no significant change of
the spectral shapes for both the sources.  Therefore, to increase
statistics, we summed the data of the two observations for the
spectral analysis as is given in Fig.~\ref{fig:sis-spec-imagefit}.
Both the spectra of AX~J131501$+$3141 and AX~J131502$+$3142 were
nicely fitted with a power-law with an absorption.  The best-fit
models and parameters are given in Fig.~\ref{fig:sis-spec-imagefit}
and Table~\ref{table:both-spec} (Methods A and B), respectively.
  Details of the analysis of the time variability are given in Sec.3.4.

% \placetable{table:both-spec}
% \placefigure{fig:sis-spec-imagefit}

\subsection{The Mixed SIS/GIS Spectrum\label{sec:sis-gis-spec}}

To confirm the above results and even to make tighter constraints on
the spectra of the two sources, we examined the spectra by an independent method.
Firstly, we accumulated SIS0$+$1 spectrum in a 3$'$ radius circle with
the center at the peak of AX~J131501$+$3141 for the summed data of the
two observations.  Since this spectrum inevitably contains photons
from both of AX~J131502$+$3142 and AX~J131501$+$3141, separated only by 1$'$,
we refer the spectrum as the ``Mixed SIS spectrum''.  We subtracted the
background taken from the region of the opposite corner in the same
SIS chip after the correction for its position dependence in the detector
plane (Ueda 1996).  We constructed Auxiliary Response Function (ARF)
at the position of AX~J131501$+$3141. The ARF at the position of
AX~J131502$+$3142 differs only by 4\%, which is negligible in the present
analysis.

\begin{figure}

%   \epsfile{file=fig/l12_s01_src3_normbgd200_hsrc_thall_rev2.ps,height=0.3\textheight}\\
%   \centering
%   \epsfile{file=fig/fig3_h.eps,height=0.4\textheight}\\
% \figcaption[fig3.eps]{
\caption[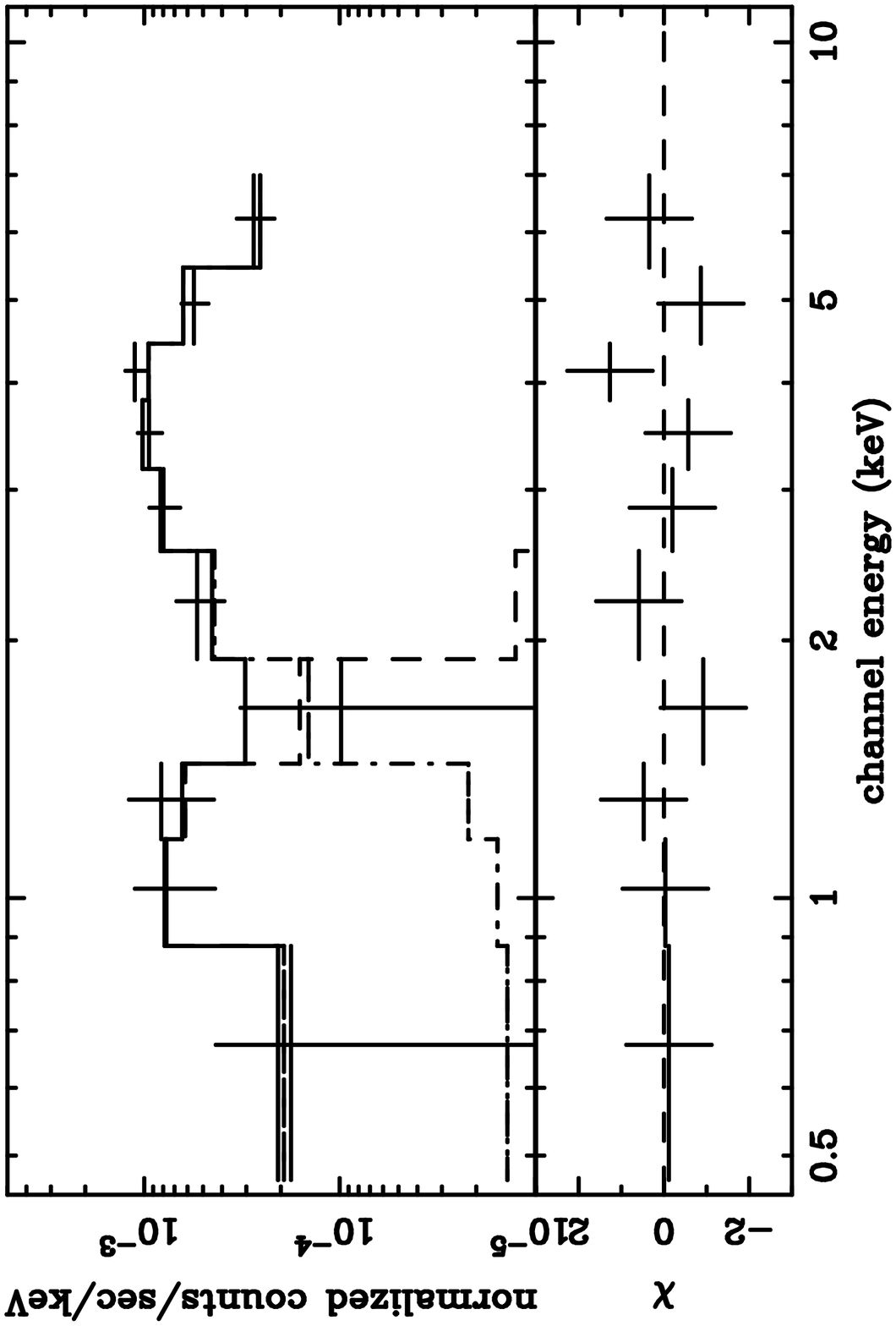]{
The mixed SIS spectrum with the best-fit of two-power-law model (Method C).
\label{fig:sis-spec-allth}
}
\end{figure}

Since we already found the presence of two power-law sources in the
image-fitted spectra, we fitted the ``Mixed SIS spectrum'' with a
model consisting of two power-laws with independent absorptions
(two-power-law model), each of which represents AX~J131502$+$3142 and
AX~J131501$+$3141. In the fitting, the power-law indices, normalizations, and column
densities for both the components are allowed to be free parameters. The
fitting result was found to be acceptable with the best-fit model and
parameters given in Fig.~\ref{fig:sis-spec-allth} and
Table~\ref{table:both-spec} (Method C). We confirmed that the results
obtained here are consistent with those obtained from the image-fitted
spectra. Note that the spectral parameters for AX~J131501$+$3141
($\Gamma=2.3^{+1.3}_{-1.0}, N_{\rm H}=7.9^{+5.3}_{-3.4}\times
10^{22}$ \NHUNIT) are more tightly constrained, while no further
constraint is given to AX~J131502$+$3142.

% \placefigure{fig:sis-spec-allth}

Since GIS has higher efficiency in the high energy band than SIS,
further constraints on the spectrum of AX~J131501$+$3141 should be given
by the GIS data. Thus, we made the GIS2$+$3 spectra from the summed
image of the two observations within a 3$'$ radius centered at the
peak of AX~J131501$+$3141. The background spectra were constructed from
a 4$'$--6$'$ radius annular region centered at the source.  The
background subtracted spectrum is given in Fig.~\ref{fig:both-fit}~(a)
(the ``Mixed GIS spectrum''). Finally, to make the tightest
constraints on the spectra, we fitted the Mixed GIS
spectrum and the Mixed SIS spectrum simultaneously with a
two-power-law model.  The fitting was acceptable within 90\%
confidence level ($\chi^2/$d.o.f $=$ 22.5/16).  The best-fit
parameters for the soft source AX~J131502$+$3142
and the hard source AX~J131501$+$3141 are $\Gamma_{\rm
soft}>-0.9$, $N_{\rm H,soft}=1.5^{+1.1}_{-0.5}\times 10^{22}$ \NHUNIT,
and $\Gamma_{\rm hard}=1.5^{+0.7}_{-0.6}$, $N_{\rm
H,hard}=6.4^{+3.1}_{-2.3}\times 10^{22}$ \NHUNIT, respectively.  The
best-fit models and parameters are given in Fig.~\ref{fig:both-fit}~(a)
and in Table~\ref{table:both-spec} (Method D).
Fig.~\ref{fig:both-fit}~(b) shows two-parameter error contours for
the photon index and the hydrogen column density of AX~J131501$+$3141,
obtained by the simultaneous fitting.

% \placefigure{fig:both-fit}

\subsection{Time Variability \label{sec:variability}}

  Since the spectral shapes of the two observations show no
significant difference (see Sec.3.2), we fixed the best-fit model
obtained with the Method D for both SIS and GIS spectra, and
separately fitted the spectrum in each observation allowing only the
normalization to vary.  The best-fit fluxes are given in
Table~\ref{table:flux}.  While the flux of AX~J131502$+$3142 did not
change significantly, that of AX~J131501$+$3141 increased by $29\pm
17$\% with a 90\% statistical error in 0.5 year from the 1st to the
2nd observation.

% \placetable{table:flux}

\section{Discussion}

We extracted the accurate spectrum of the LSS hardest source,
AX~J131501$+$3141, taking account of the contamination from the nearby
soft source, AX~J131502$+$3142, from which no significant X-ray emission
was found in the LSS (Ueda 1996).
 We found that AX~J131501$+$3141 exhibits a
large absorption of $N_H = (6^{+4}_{-2})\times 10^{22}$ \NHUNIT\ with
a photon index $\Gamma = 1.5^{+0.7}_{-0.6}$. It showed a long-term time
variability between two observations separated by 0.5 year.
 While the photon index of AX~J131501$+$3141
is consistent with the canonical value of type 1 AGNs (e.g., Mushotzky 1993),  
its absorption column density is larger than that of
typical type 1 AGN by more than an order of magnitude, 
although a part of type 1 AGNs, about 10\% of them (\cite{Schartel97}),
shows a column density larger than $5\times 10^{22}$ \NHUNIT.

It is important that this source is selected fully unbiasedly.  Hence,
its X-ray properties should provide a key to understand the general
nature of the missing hard X-ray populations which constitute the CXB
above 2 keV. Two major possibilities have been proposed to account
for the apparent hard spectrum of the CXB: one is to introduce large
absorptions of sources (e.g., \cite{Awaki91}), and the other is to
consider populations of sources with intrinsically flat spectra (e.g.,
\cite{Morisawa90}; \cite{Matteo97}). Our results of the LSS
hardest source strongly suggest that highly absorbed sources play an
important role in considering the origin of the hard X-ray background.

The large absorption of $6\times 10^{22}$ \NHUNIT, the photon index of
$\Gamma\sim 1.5$, and the time variability are common properties seen in
type 2 Seyfert galaxies.  In fact, systematic studies of type 2 Seyfert
galaxies by Awaki \etal\ (1991) and Ueno (1996) revealed that they
commonly show large absorptions of $\sim 10^{23}$ \NHUNIT\ and photon
indices of 1.5--1.7. Akiyama \etal\ (1998) found one bright optical
galaxy with $B=17.25$ mag near the center of the X-ray error circle of
0.5$'$ radius in the optical follow-up observations. No other optical
source with the flux larger than $B=22.4$ mag is found in the error
circle. Akiyama \etal\ (1998) performed spectroscopic observations of
the bright galaxy and found that ratios of emission lines are similar
to those found in type 2 Seyfert galaxies. The redshift of this galaxy
was determined to be 0.07. From the redshift, the observed flux in the
2--10 keV band, $5\times 10^{-13}$ \FLUXUNIT, can be converted to the
absorption corrected luminosity of $L_{\rm X}\sim 2\times 10^{43}\
{\rm erg~s}^{-1}$.  This luminosity is consistent with those of
Seyfert galaxies. Thus, we identify AX~J131501$+$3141 found in
the unbiased X-ray survey as a type 2 Seyfert galaxy.
  Using the LogN-LogS relation in Hasinger \etal\ (1998), we estimate that the chance
coincidence between AX J131501$+$3141 and AX J131502$+$3142 is $\sim$ 3\%.
However, these two sources have probably no physical correlation,
 because AX J131502$+$3142 is likely to be a QSO\footnote{
 In the optical imaging observations of R- and B-band by Akiyama \etal\ (1998),
we found two point-like optical sources
located at about 10$''$ north from the center of the error circle with 30$''$
radius for the soft source (AX J131502$+$3142).
They are very close to each other and their magnitudes are comparable.
Total magnitudes of the two sources are B=20.8 mag and R=19.6 mag; B-R color
is 1.21.
Since both the optical color and the optical to soft X-ray flux
 is consistent with that for type-1 AGNs,
it is possible that one of them is a quasar which is also responsible for
AX J131502$+$3142.
However there are fainter optical sources (R$\gtrsim$22) in the error circle
 and the optical identification for AX J131502$+$3142 is not clear yet.
}
 which is more distant
 than the new type 2 Seyfert AX J131501$+$3141.

Awaki (1991), Madau, Ghisellini \& Fabian (1994) and Comastri \etal\
(1995) predicted that the combination of type 1 and type 2 AGNs
can reproduce the CXB spectrum, based on the unified AGN scheme
(e.g., \cite{Antonucci93}).  In the scheme, type 1 and type 2 AGNs are
essentially the same objects, observed from different
viewing angle. These type 2 AGNs, which exhibit apparently fainter
and harder X-ray spectra than those of type 1 AGNs, should become
detectable as the detector sensitivity increases. Although we have
examined only one sample from the LSS at this moment, the result is
encouraging not only for the unified AGN scheme, but also for solving
the origin of the CXB.

\acknowledgments

    The authors express their thanks all of the members of the \ASCA\ team
 whose efforts made these observations and data analysis possible.
  We are grateful to H.~Inoue for his valuable comments.
  We thank the referee, G.~Hasinger, for his useful advice.
  M. S. thanks Y.~Maeda for discussion.  
  M. S. and M. A. acknowledge the supports from
 the Japan Society for the Promotion of Science for Young Scientists.

\newpage

\begin{figure}
%   \epsfile{file=fig/L12s01_soft_3bin.ps,width=\textwidth}\\
%   \epsfile{file=fig/L12s01_hard_3bin.ps,width=\textwidth}\\
%  \centering
%  \epsfile{file=fig/fig2a_h.eps,height=0.4\textheight}\\
%  \epsfile{file=fig/fig2b_h.eps,height=0.4\textheight}\\
% \figcaption[fig2.eps]{
\caption[fig2.eps]{
The SIS0$+$SIS1 spectra obtained with the image-fit.
Both spectra are fitted with an absorbed power-law (Method A, B).
(a) AX~J131502$+$3142 (Soft source)
(b) AX~J131501$+$3141 (Hard source)
\label{fig:sis-spec-imagefit}
}
\end{figure}

\begin{figure}
%   \epsfile{file=fig/l12all_matomeg120_th_rev_3.ps,height=0.3\textheight}\\
%   \centering
%   \epsfile{file=fig/fig4a_h.eps,height=0.4\textheight}\\
%   \epsfile{file=fig/fig4b_h.eps,height=0.4\textheight}\\
% \figcaption[fig4.eps]{\protect
\caption[fig4.eps]{
(a) Same as Fig.\ref{fig:sis-spec-allth} but for the GIS spectrum
 and the best-fit model in the simultaneous fitting (Method D) of SIS$+$GIS data.
(b) The 68, 90, 99\% confidence contours for the photon index and hydrogen column density
of AX~J131501$+$3141 in the simultaneous fitting of SIS$+$GIS data.
\label{fig:both-fit}
}
\end{figure}

\clearpage

\begin{table}
\begin{center}
\caption[]{The fitting results
\label{table:both-spec}
}
\begin{tabular}{lccccc}
\tableline
\tableline
 & \multicolumn{2}{c}{AX~J131502$+$3142} & \multicolumn{2}{c}{AX~J131501$+$3141} & \\
 & \multicolumn{2}{c}{(Soft component)} & \multicolumn{2}{c}{(Hard component)} & \\
Method\tablenotemark{a} & Photon index & $N_H$ & Photon index & $N_H$ & $\chi^2/d.o.f.$ \\
 &  & ($10^{22}$ \NHUNIT) & & ($10^{22}$ \NHUNIT) & \\
\tableline
(A) & $3.8^{+>6}_{-2.6}$ & $0.7\pm 2.4$ & \nodata & \nodata & 2.5/3 \\
(B) & \nodata & \nodata & $2.8^{+2.7}_{-1.8}$ & $11^{+13}_{-6}$ & 2.6/3 \\
(C) & $>1.3$ & $1.7^{+1.0}_{-0.6}$ & $2.3^{+1.3}_{-1.0}$ & $7.9^{+5.3}_{-3.4}$ & 4.3/4 \\
(D) & $>-0.9$ & $1.5^{+1.1}_{-0.5}$ & $1.5^{+0.7}_{-0.6}$ & $6.4^{+3.1}_{-2.3}$ & 22.5/16 \\
\tableline
\end{tabular}
\end{center}
\scriptsize
\noindent
\tablenotetext{a}{Description of the methods:\\
\begin{tabular}{ll}
(A)& The SIS0$+$1 spectrum for the soft source AX~J131502$+$3142 made with image-fitting.\\
   & An absorbed power-law with all free parameters is used.\\
(B)&  Same as (A), but for the hard source AX~J131501$+$3141.\\
(C)&  The mixed SIS Spectrum fitted with a two-component power-law model with absorption. \\
(D)&  Same as the model (C) but for  a combined fitting with SIS and GIS.\\
\end{tabular}
}
\end{table}

\clearpage

\begin{table}
\begin{center}
\caption[]{The fluxes of AX~J131502$+$3142 and AX~J131501$+$3141 
\label{table:flux}}
\begin{tabular}{llccc}
\tableline
\tableline
                 &           &        & 1st epoch. & 2nd epoch. \\
\tableline
AX~J131502$+$3142&(0.5--2 keV)& SIS    & $0.16\pm 0.08$ & $0.23\pm 0.08$ \\
\tableline
AX~J131501$+$3141&(2--10 keV) & SIS    & $5.1\pm 0.8$ & $6.7\pm 0.8$\\
                 &           & GIS    & $4.5\pm 0.7$ & $5.7\pm 0.7$ \\
                 &           &SIS$+$GIS\tablenotemark{a}& $4.8\pm 0.5$ & $6.2\pm 0.5$ \\
\tableline
\end{tabular}
\end{center}
\scriptsize
\tablecomments{Unit of flux is $10^{-13}$ \FLUXUNIT. Each error is 90\% confidence.}
\tablenotetext{a}{Mean flux of AX~J131501$+$3141 about SIS and GIS.}
\end{table}


\clearpage

\begin{thebibliography}{}

\bibitem[Akiyama et al. 1998]{Akiyama98}
Akiyama, M., \etal\ 1998, \apj, in press, astro-ph/9801173

\bibitem[Antonucci 1993]{Antonucci93}
Antonucci, R. 1993, \araa, 31, 473

\bibitem[Awaki 1991]{Awaki91}
Awaki, H. 1991, Ph.D. thesis, Nagoya Univ.

\bibitem[Awaki et al. 1991]{Awakietal91}
Awaki, H., Koyama, K., Inoue, H., \& Halpern, J. P. 1991, \pasj, 43, 195

\bibitem[Burke et al. 1991]{Burke91}
Burke, B. E., Mountain, R. W., Harrison, D. C., Bautz, M. W., Doty, J. P., Ricker, G. R., \& Daniels, P. J. 1991, IEEE Trans. ED-38, 1069

\bibitem[Burke et al. 1994]{Burke94}
Burke, B. E., Mountain, R. W., Daniels, P. J., \& Dolat, V. S. 1994, IEEE Trans. Nuc. Sci., 41, 375

\bibitem[Comastri et al. 1995]{Comastri95}
Comastri, A., Setti, G., Zamorani, G., \& Hasinger, G. 1995, \aap, 296, 1

\bibitem[Day et al. 1995]{Day95}
Day, C., Arnaud, K., Ebisawa, K., Gotthelf, E., Ingham, J., Mukai, K., \& White, N. 1995, The ABC Guide to \ASCA\ Data Reduction, NASA/Goddard Space Flight Center

\bibitem[Fabian \& Barcons 1992]{Fabian92}
Fabian, A. C., \& Barcons, X. 1992, \araa, 30, 429

\bibitem[Gendreau et al. 1995]{Gendreau95}
Gendreau, K. C., et al. 1995, \pasj, 47, L5

\bibitem[Giacconi et al. 1962]{Giacconi62}
Giacconi, R., Gursky, H., Paolini, F. R., \& Rossi, B. B. 1962, \prl, 9, 439

\bibitem[Hasinger 1996]{Hasinger96}
Hasinger, G. 1996, \aaps, 120, 607

\bibitem[Hasinger et al. 1998]{Hasinger98}
Hasinger, G., Burg, R., Giacconi, R., Schmidt, M., Tr\"umper, J., \& Zamorani, G. 1998, \aap, 329, 482

\bibitem[Inoue et al. 1996]{Inoue96}
Inoue, H., Kii, T., Ogasaka, Y., Takahashi, T., \& Ueda, Y. 1996, in proc. of ``R\"ontgenstrahlung from the Universe'', eds. Zimmermann, H. U., Tr\"umper, J., \& Yorke, H., p323

\bibitem[Kunieda et al. 1995]{Kunieda95}
Kunieda, H., Furuzawa, A., Watanabe, M., \& the \ASCA\ XRT Team 1995, the \ASCA\ News No.3, NASA/Goddard Space Flight Center, p3

\bibitem[Madau et al. 1994]{Madau94}
Madau, P., Ghisellini, G., \& Fabian, A. C. 1994, \mnras, 270, L17

\bibitem[Makishima et al. 1996]{Makishima96}
Makishima, K., \etal\ 1996, \pasj, 48, 171

\bibitem[Di Matteo \& Fabian 1997]{Matteo97}
Di Matteo, T., \& Fabian, A. C. 1997, \mnras, 286, 393

\bibitem[McHardy et al. 1998]{McHardy98}
McHardy, I. M., \etal\ 1998, \mnras, in press

\bibitem[Morisawa et al. 1990]{Morisawa90}
Morisawa, K., Matsuoka, M., Takahara, F., \& Piro, L. 1990, \aap, 236, 299

\bibitem[Mushotzky 1993]{Mushotzky93}
Mushotzky, R. F. 1993, \araa, 31, 717

\bibitem[Ohashi et al. 1996]{Ohashi96}
Ohashi, T., \etal\ 1996, \pasj, 48, 157

\bibitem[Serlemitsos et al. 1995]{Serlemitsos95}
Serlemitsos, P. J., \etal\ 1995, \pasj, 47, 105

\bibitem[Schartel et al. 1997]{Schartel97}
Schartel, N., Schmidt, M., Fink, H. H., Hasinger, G., \& Tr\"umper, J. 1997, \aap, 320, 696

\bibitem[Tanaka, Inoue, \& Holt 1994]{Tanaka94}
Tanaka, Y., Inoue, H., \& Holt, S. S. 1994, \pasj, 46, L37

\bibitem[Ueda 1996]{Ueda96}
Ueda, Y. 1996, Ph.D. thesis, Tokyo Univ.

\bibitem[Ueda et al. 1998a]{Ueda98a}
Ueda, Y., \etal\ 1998a, \nat, 391, 866
%Ueda, Y., \etal\ 1998a, \nat, in press

\bibitem[Ueda et al. 1998b]{Ueda98b}
Ueda, Y., \etal\ 1998b, will appear in \ASCA\ News, NASA/Goddard Space Flight Center

\bibitem[Ueno 1996]{Ueno96}
Ueno, S. 1996, Ph.D. thesis, Kyoto Univ.

\bibitem[Uno 1997]{Uno97}
Uno, S. 1997, Ph.D. thesis, Gakusyuin Univ.

\bibitem[Vikhlinin et al. 1995]{Vikhlinin95}
Vikhlinin, A., Forman, W., Jones, C., \& Murray, S. 1995, \apj, 451, 553

\end{thebibliography}
\end{document}